\documentclass{revtex4-2}
\usepackage{graphicx}
\usepackage{dcolumn}
\usepackage{bm}

\begin{document}

\title{Computational Investigations of Selected Enzymes From Two Iron and $\alpha$--ketoglutarate--Dependent Families}

\author{Madison B. Berger}
 \altaffiliation[]{Chemistry Department, University of North Texas.}
\author{Alice R. Walker}%
 \altaffiliation[]{Chemistry Department, Wayne State University.}
\author{Erik Antonio V\'azquez Montelongo}
 \altaffiliation[]{Chemistry Department, University of North Texas.}
\author{G. Andr\'es Cisneros}
 \altaffiliation[]{Chemistry Department, University of North Texas.}

\date{\today}

\begin{abstract}
DNA alkylation is used 
as the key epigenetic mark in eukaryotes, however, most
alkylation in DNA can result in deleterious effects. Therefore, this
process needs to be tightly regulated.
AlkB and TET are families within the Fe and $\alpha$-kg-dependent
superfamily of enzymes that are tasked with dealkylating DNA and RNA in cells. Members of these 
families span all species and are an integral part of transcriptional regulation. 
While both families catalyze oxidative dealkylation of various bases, 
each has specific preference for alkylated base type as
well as distinct catalytic mechanisms. This perspective aims to provide 
an overview of computational work carried out to investigate
several members of these enzyme families
including AlkB, ALKBH2, ALKBH3 and TET2. Insights
into structural details, mutagenesis studies, reaction path analysis, 
electronic structure features in the active site, and 
substrate preferences are presented and discussed
\end{abstract}

\maketitle

\section{Introduction}

The Fe/$\alpha$--ketoglutarate (Fe/$\alpha$-kg) dependent 
superfamily of enzymes encompasses a large number of enzyme 
families that catalyze a broad range of reactions. \cite{Hausinger0421,Martinez1520702,Herr18517,Islam18585,Yu21}
While there are many enzymes very similar to those that make up this family in sequence 
and/or mechanism, no others require the substrate $\alpha$-kg. These enzymes are known to act
on amino acid side chains, lipids, nucleotides and a variety of small molecules.
\cite{Hausinger0421}
The reactions carried out include hydroxylation, demethylation, epoxidation, ring fragmentation, 
epimerization and desaturation. \cite{Gao18792,Islam18585,Hausinger151}
Several Fe/$\alpha$-kg enzymes and their reaction mechanisms have been investigated by various
computational approaches.
\cite{Alvarez-Barcia175347,Hibi143869,Wang1413895,Goudarzi205152,Zwick201065,Wu16453,Loenarz117,Sun181,Mandal18107} 
Among this superfamily, members that are involved in DNA
transactions have been a particular focus. 

DNA is susceptible to various types of damage and modifications that
frequently arise due to endogenous and/or exogenous sources.\cite{Branzei08297}  Alkylation is a
common form of DNA modification and can have both negative and positive effects. Methylation of cytosine at the 5 position to
yield 5-methylcytosine is the staple epigenetic marker for gene
regulation \cite{Kumar2018640}. 
Conversely, other types of alkylation can lead to instabilities and strand breakage, which
has been exploited as a common form of cancer treatment .\cite{Drablos041389}
Two of the enzyme families tasked with handling this type of damage 
or modification are the AlkB and Ten-Eleven
Translocation (TET) families. \cite{Fedeles1520734,Leddin1991} Both 
of these families are members of the Fe/$\alpha$-kg superfamily.\cite{Zheng144602, Hausinger0421}
These enzymes have been extensively studied using computational, experimental, 
and crystallographic methods because of their biochemical importance. \cite{Hausinger0421} 
AlkB of \textit{E.coli} 
is responsible for the oxidative dealkylation of nucleobases. 
These enzymes can act on 
all seven of the N-methylated Watson-Crick base pairs such as N\textsuperscript{1}-
methyl adenine (1mA) and N\textsuperscript{3}-methyl cytosine (3mC). \cite{Chen16687} 
Various homologues
of \textit{E. coli} AlkB exist across both prokaryotes and eukaryotes. \cite{Fedeles1520734}

DNA methylation is a key epigenetic modification that is conducted 
by the DNA methyltransferase family. \cite{Lyko1881} 
A balance between methylation and demethylation
must be met in order to preserve the integrity of the genome. 
Disruptions to the demethylation of cytosine
is common in many types of cancer. \cite{Cheishvili152705} 
TET enzymes catalyze the sequential oxidation of 
5-methylcytosine (5mC) to 5-hydroxymethylcytosine (5hmC), 5-formylcytosine (5fC) and
5-carboxylcytosine (5caC)
(Fig.\ref{fgr:tet2-mech}B).\cite{Ito111300}
The TET family consists of three members all located on 
human chromosomes: TET1, TET2 and TET3. \cite{Li15171}
All TET enzymes are a crucial component of tumor suppressor 
mechanisms in cancer and it is well known 
that all three are easily mutated. \cite{Rasmussen16733} 
Damage to TET1 can lead to acute 
myeloid and lymphoid leukemias
and TET2 damage can lead to several different types of myeloproliferative disorders.
\cite{Tsiouplis21}
 
\begin{figure}
 \centering
 \includegraphics[height=4.45cm]{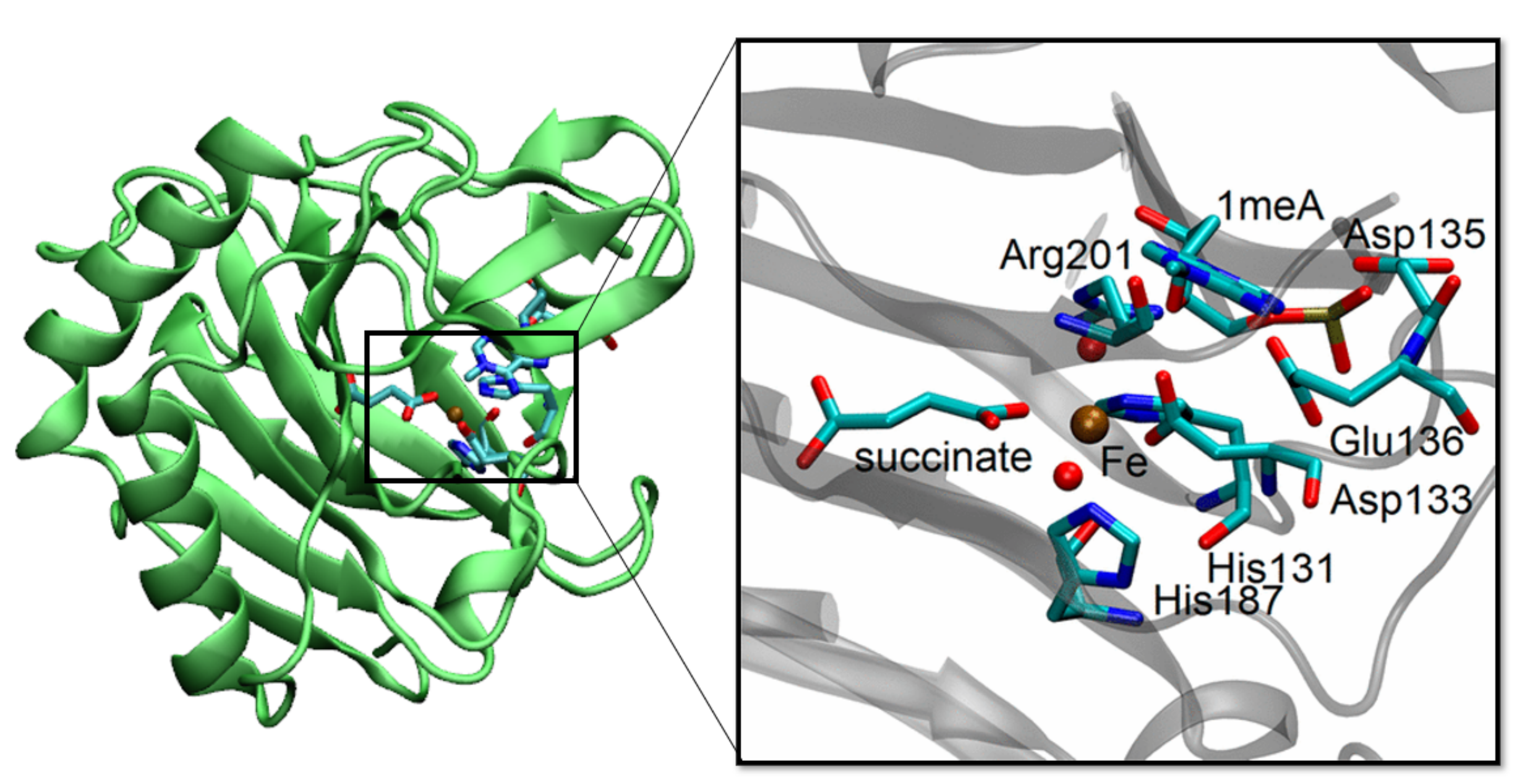}
 \caption{The active site of AlkB containing succinate. Protein structure generated from pdb: 
 2FDG.\cite{Edstrom06879} Active site image reproduced from: D. Fang, G.A. Cisneros, 
 \textit{J. Chem. Theory Comput.}, 2014, \textbf{10}, 5136--5148. 
 (doi/10.1021/ct500572t) Copyright 2014 American Chemical Society.}
 \label{fgr:alkb-active-site}
\end{figure}

The reaction mechanisms for both AlkB and TET were 
initially inferred from TauD, another member of the
Fe/$\alpha$-kg dependent family. \cite{Hausinger0421,Usharani11176} 
There are two features that are structurally conserved 
in this family. A non--heme Fe, which alternates between 
oxidation states II, III and IV depending on the stage of the 
reaction. The non--heme Fe is coordinated by a triad of amino acids 
in the active site consisting of two histidine (His) residues and 
one aspartate (Asp) or glutamate (Glu) (Fig.\ref{fgr:alkb-active-site} and Fig. \ref{fgr:tet2-active-site}).
\cite{Liu094887} 
This feature is also referred to as the 2-His-1-carboxylate triad. \cite{Hegg97625}
In addition, this triad is located inside of a double-stranded $\beta$-sheet fold,
also known as a "jelly-roll" fold. The Fe maintains an octahedral geometry with 
three of those sites 
occupied by the 2-His-1-carboxylate triad and the other three by water 
or the required co--substrates: $\alpha$--ketoglutarate and
molecular oxygen. 

Members from both of these enzyme families have been investigated
by both classical and hybrid quantum mechanics/molecular mechanics
(QM/MM) approaches to gain insights into their structure, function
and catalytic mechanisms. Molecular dynamics (MD) simulations 
have been employed to study the structural features of AlkB and 
TET2, including the effects of mutations and different types 
of RNA/DNA.
\cite{DeNizio19411,Liu17181,Waheed213877,Walker171,Silvestrov14123,Lenz201}

Quantum mechanics/molecular mechanics (QM/MM) is a useful
method to study the reactivity of enzymes. Briefly, 
a small number of atoms in the system are treated with 
DFT, semi-empirical or other {\em ab initio} approaches
and the rest of the protein is treated with (generally) a classical
force field. \cite{Senn091198}
One of the many challenges of QM/MM can be separating a system 
into the two regions and determining which
residues are important for the reaction of interest. Several methods 
currently suggested in the literature for building
the QM region include protein sequencing and structural evolution analyses, 
electron localization function
or charge shift analysis. \cite{Hix21,Kulik1611381} Once the regions have been determined, 
the boundary between the two must be specially handled, especially if there are bonds 
that are "cut" across the QM/MM boundary.
There are a variety of methods used throughout the literature 
to address this issue including the pseudobond approach, link atoms and frozen localized orbitals.\cite{Senn091198}

Another key component of QM/MM is the type of charge embedding that will be used between the 
QM charge density and the force field used in the MM. Three separate schemes have been designed
to better represent the interactions between the two regions: mechanical,
electrostatic, and polarizable embedding. \cite{Dohn201} Mechanical embedding is the 
most basic method for
representing interactions in that a force field is used for the QM and MM interaction. 
Electrostatic embedding is the most popular method and not as 
computationally demanding as polarizable embedding. Here, the charges from the MM
are incorporated into the calculation of the QM Hamiltonian. 
Polarizable embedding takes electrostatic
embedding a step further and allows for self consistent polarization
between the quantum and classical subsystems. \cite{Nottoli20,Nochebuena2021e1515,Loco20212812} 

Once the QM/MM system has been established, it can be used to 
compute properties and investigate
reaction pathways. There are several methods that can then be used to determine the minimum 
energy paths for chemical reactions, including a family of methods
called chain-of-states. This family includes approaches such as the
Nudged Elastic Band, replica path, and the quadratic string methods.\cite{Zarkevich15,Woodcock03140,Burger06} 
In these chain-of-states methods, appropriate sampling of the 
various states of the system must be met
in order to connect the first and last state (reactant and product). 
QM/MM free energy calculations can also provide insight on 
reaction barriers, solubility, substrate binding affinity
and equilibrium constants. \cite{Hansen142632,Lu171056} 
One approach to compute the free energy associated with
the reactive process in an enzyme is called the minimum free energy path. \cite{Hu07390,Hu08} 

In this perspective, we discuss how computational simulations
have been used to elucidate the reaction mechanisms of selected 
AlkB and TET family enzymes. We show how combining QM/MM with 
other tools and techniques such as MD, non-covalent interactions,
and other analyses continue to provide important insights into the
reactivity and function of these important systems.  
The remainder of the paper is as follows: Section II 
presents the computational investigation of selected AlkB family
enzymes, namely AlkB, ALKBH2 and ALKBH3. This section is separated
following the different stages of the enzymatic reaction mechanism,
and provides new QM/MM results on the rate limiting step catalyzed
by ALKBH2 and ALKBH3. Subsequently, computational investigation
of the mechanism of TET2 are presented in Section III,
again divided by the different oxidation stages catalyzed by
this enzyme. Finally, concluding remarks and and a perspective
are presented.
\section{Mechanistic Studies of AlkB Family Enzymes}
\label{sectalkb}

\begin{figure*}
 \centering
 \includegraphics[height=14cm]{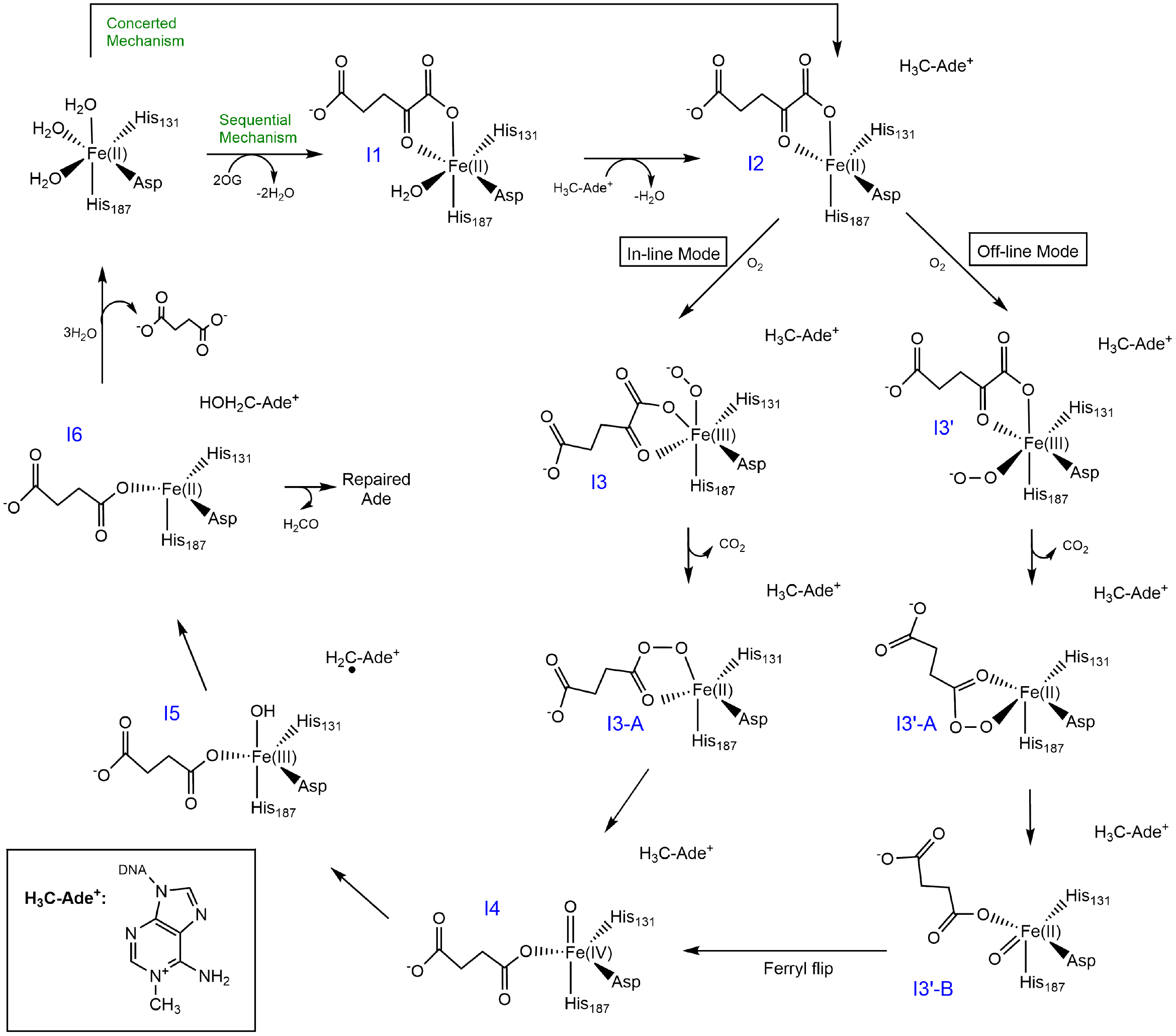}
  \caption{The proposed AlkB catalyzed dealkylation mechanism of 1mA 
 adapted from \textit{Waheed et al}.\cite{Waheed20795}}
 \label{fgr:alkb-mech}
\end{figure*}

The AlkB family of enzymes have the ability to repair both monoalkyl substrates as well as certain 
etheno adducts of some DNA and RNA bases, with AlkB being the most versatile. 
The repair mechanism for both types of damage are very similar. 
\cite{Delaney0414051,Koivisto0440470,Li101,Li128896,Li131182,Delaney05855,Maciejewska1024,Mishina0514594,Duncan0216660,Zdzalik151}
The reaction mechanism can be separated in two phases and is summarized in Figure
\ref{fgr:alkb-mech}. The first phase
involves the formation of a reactive ferryl, Fe(IV)=O, intermediate;
this is then followed by the oxidation of the alkyl moiety on the
substrate. This dealkylation reaction carried out by the AlkB 
family of enzymes has been extensively studied via 
MD, QM/MM and DFT methods.

Using the repair of 1mA as an example, the first step 
is the formation of the ferryl intermediate via 
binding of $\alpha$-kg and O\textsubscript{2}. This results in the formation 
of succinate and CO\textsubscript{2}. From there, the oxyl atom on the Fe(IV)=O
intermediate must rearrange from an axial to an equatorial 
position in order to align the oxygen trans to a neighboring His residue.
\cite{Liu094887,Sastri0719181}
Subsequently, an H atom is abstracted by the oxyl 
from the substrate (1mA), which is also the rate limiting 
step of this reaction followed by an OH rebound to the
radical methyl moiety. Finally, 1-hydroxymethyladenine 
disproportionates into formaldehyde and the 
dealkylated base.

\subsection{Substrate Binding}

Before the coordination of $\alpha$-kg and O\textsubscript{2} to the
Fe(II) atom in the active site, three coordination sites 
of the Fe are occupied by water (and the other three are bound to
the three conserved active site residues). \cite{Martinez1520702} 
In the first phase of the reaction, $\alpha$-kg and the methylated substrate
must bind in the active site. The binding of $\alpha$-kg displaces two of the equatorial waters
allowing for a bidentate configuration with the Fe (\textbf{I1}). 

Both sequential and concerted mechanisms
have been proposed for the binding of $\alpha$-kg and 1mA. \cite{Solomon21} Magnetic circular
dichroism spectrosopy results show that the binding of $\alpha$-kg alone does not allow for a rapid
formation of Fe(IV)=O. Both substrates must be simultaneously bound 
in order for the dioxygen activation and substrate oxidation to continue which suggests 
that a concerted mechanism is preferred. 
Given that the concerted mechanism has been shown to be the most likely, the question
can then be raised as to whether the O\textsubscript{2} can even enter the active
site let alone bind to the Fe center without the other two substrates binding first. 
The authors also noted the penta-coordinate geometry around the 
Fe before O\textsubscript{2} binding. The binding of both substrates is what 
allows for the opening of a position on the Fe(II) that is meant for O\textsubscript{2}.
In a similar system to AlkB where pterin was used instead of $\alpha$-kg, the calculated barrier 
for product release was 16 kcal/mol. If the pterin was not bound, this barrier increases
by 5 kcal/mol, resulting in a much slower turnover rate by the enzyme. 

\begin{figure}
 \centering
 \includegraphics[height=7cm]{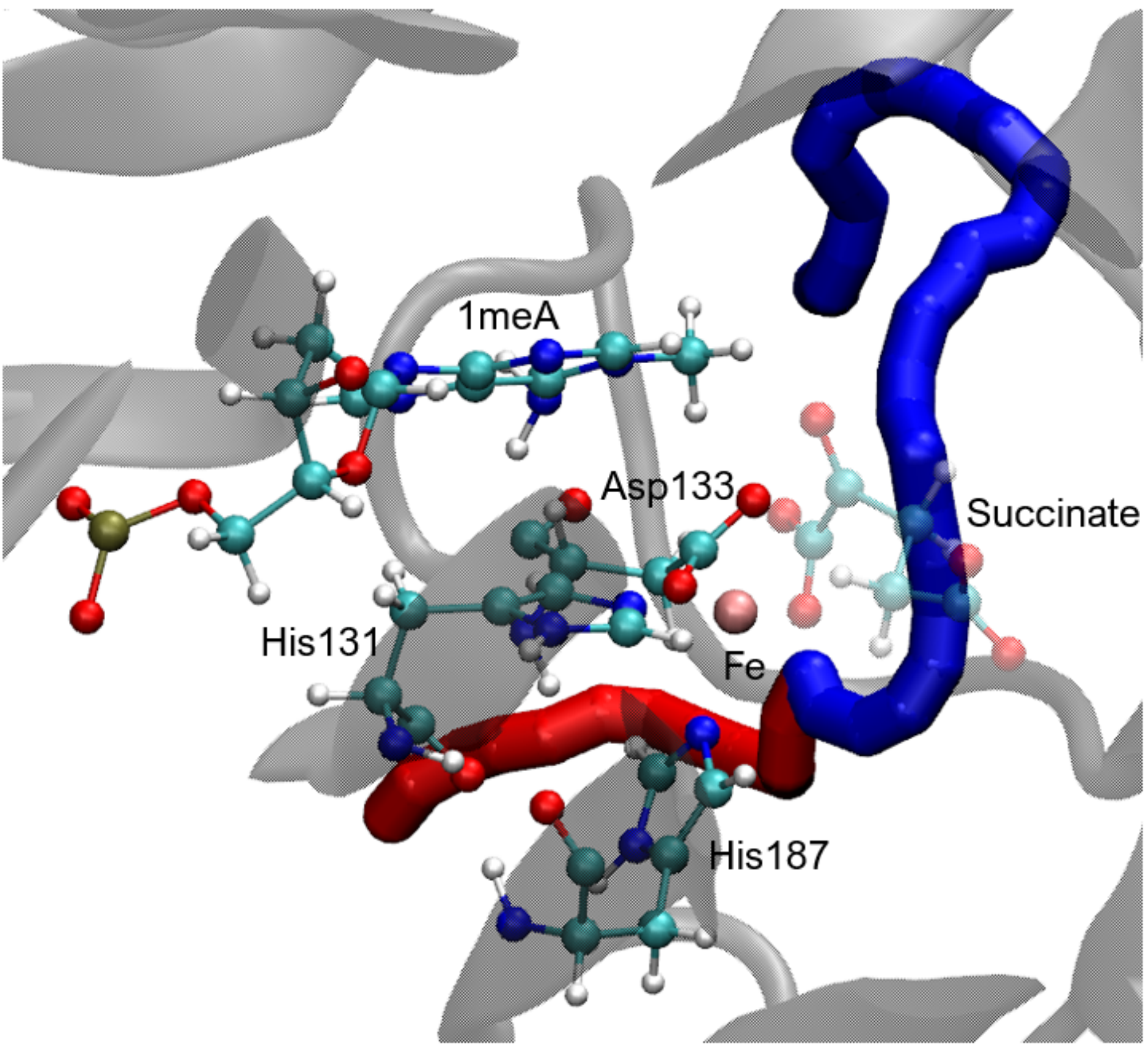}
 \caption{The two most probable tunnels for O\textsubscript{2} to travel in order to reach the
 active site adapted from Torabifard \textit{et al.}\cite{Torabifard176230}}
 \label{fgr:alkb-tunnels}
\end{figure}

Subsequently, O\textsubscript{2} must replace the third water on the Fe coordination sphere and 
complete the octahedral coordination to the Fe.
The mechanism for the binding of the O\textsubscript{2} co--substrate
to complete the enzyme--substrate (ES) complex was hypothesized 
to occur via an intra--molecular tunnel observed in the original 
crystal structures. \cite{Yu06879} 
The proposed O\textsubscript{2} transport mechanism was recently 
investigated using polarizable and non--polarizable MD
simulations paired with several other analyses (Figure \ref{fgr:alkb-tunnels}).
\cite{Torabifard176230} 
Computational analysis suggested two possible tunnels by which the 
O\textsubscript{2} might transfer from the solvent to the 
active site. \cite{Pavelka16505,Torabifard176230} 

Based on the original tunnel observed from the crystal structure,
a tyrosine at position 178 was hypothesized to act as a possible
gate along the tunnel. \cite{Yu06879}
This tyrosine was computationally investigated via MD by replacing it
with three different amino acids (W178Y, W178A and W178P). 
The computational results suggest that that W178Y and W178P behave
similar to WT in O\textsubscript{2} occupancy, RMSF and distance 
correlation analyses. W178A differed
the most in dynamic behavior and revealed a new path in which the 
O\textsubscript{2} molecule could be transported.

The PMF associated with the
diffusion of O\textsubscript{2} from the surface of the protein
to the active site was calculated using the ff99SB force field 
for both possible tunnels. The most probable tunnel (blue) shown in Fig. \ref{fgr:alkb-tunnels} 
exhibited a downhill free energy barrier of 3 kcal/mol
compared to 1.5 kcal/mol (red). \cite{Torabifard176230}
Since O\textsubscript{2} is neutral, but has a large polarizability,
the barrier for the second tunnel was also calculated
using the AMEOBA polarizable force field, resulting in a calculated
free energy difference of 51.5 kcal/mol.
This large discrepancy between the non--polarizable and polarizable
PMFs is due exclusively to the polarization interactions.
The large free energy change calculated with AMOEBA is also
consistent with physiological expectations given the fact that
AlkB and homologues are adaptive-response proteins
localized mainly in the nuclei, where O\textsubscript{2} 
concentration needs to be tightly regulated to avoid DNA oxidative. Thus, high affinity
for this substrate should be expected to enable catalytic turnover.

There are two possible orientations that have been proposed for the 
binding of O\textsubscript{2} to the Fe center, termed "in-line" 
and "off-line" modes (Figure \ref{fgr:alkb-mech}). \cite{Waheed20795,Quesne14435} 
The "in-line" mode refers to the rearrangement of $\alpha$-kg around the Fe that 
allows for the opening of a coordination site that the O\textsubscript{2} can occupy.
From there, the O\textsubscript{2} binds trans to His187 (\textbf{I3}).
Conversely, the "off-line" mode corresponds to the binding of O\textsubscript{2} 
trans to His131 (\textbf{I3'}).
This path requires a ferryl flip (\textbf{I3'-B} to \textbf{I4}) in order to
properly orient the oxyl moiety so that it is closer to the 
methylated substrated. 

Several crystallographic studies have shown that the O\textsubscript{2} should 
bind in the same
position that the final water occupies before leaving the active site (trans to His131).
\cite{Yi12671,Yang08961,Yu06879} 
An MM model was used to further analyze the binding mode of O\textsubscript{2}.
\cite{Quesne14435} Quesne \textit{et al.} found that there was sufficient space 
in the binding pocket trans to His131 ("off-line" mode) to 
insert O\textsubscript{2}. An additional MM model was constructed in which the 
O\textsubscript{2} was placed trans to His187 ("in-line" mode). This mode failed to
optimized because the binding pocket was too dense at that position.
Close contacts (\textless 1.7\AA) to the Asp133 and methyl group of the substrate were found in
addition to clashes between the the O\textsubscript{2} and $\alpha$-kg.
The O\textsubscript{2} was found to be almost 6 \AA \ from the methyl carbon of 
the substrate where the proton will be abstracted. \cite{Liu094887} This would 
indicate that the proposed "in-line" mode of binding is 
the preferred pathway. 
The end point for the two tunnels proposed by Torabifard \textit{et al.}
\cite{Torabifard176230} suggest preferential binding trans to His131 (\textbf{I3'}), 
although there appears
to be sufficient space for O\textsubscript{2} to  bind trans 
to His187 and thus both binding modes appear to be accessible 
at the end of the tunnels (Figure \ref{fgr:alkb-tunnels}, \textbf{I3}). 

\subsection{Reorientation of the oxo moiety}

Upon release of CO\textsubscript{2}, the oxo moiety of Fe(IV)=O 
must undergo a reorientation from an axial 
to an equatorial position if it was added via the "in-line" mode. 
This will convert it from trans with 
respect to (w.r.t.) His131 to trans w.r.t. His187. Quesne \textit{et al.} used 
QM/MM to study the reorientation of 
the oxo moiety in order to determine why it may be catalytically relevant 
for this enzyme.\cite{Quesne14435} 
They found that the rotation of the oxo group changes the shape of the molecular
orbitals in the active site as well as the occupancies of the HOMO and LUMO orbitals.
\cite{Quesne14435} 
In the axial position, $\sigma^*_{x^2-y^2}$ is singly occupied and $\sigma^*_{z^2}$ is virtual. 
The ordering of the orbitals is completely reversed when the Fe(IV)=O reorients 
and the oxo is in the equatorial position. 
Additionally, an isomerization
energy difference of -6.0 kcal/mol in favor of the equatorial position was reported. 
It was also noted that there were very 
few changes in spin densities and charges of these two complexes. \cite{Quesne14435}

\begin{figure*}
 \centering
 \includegraphics[height=6cm]{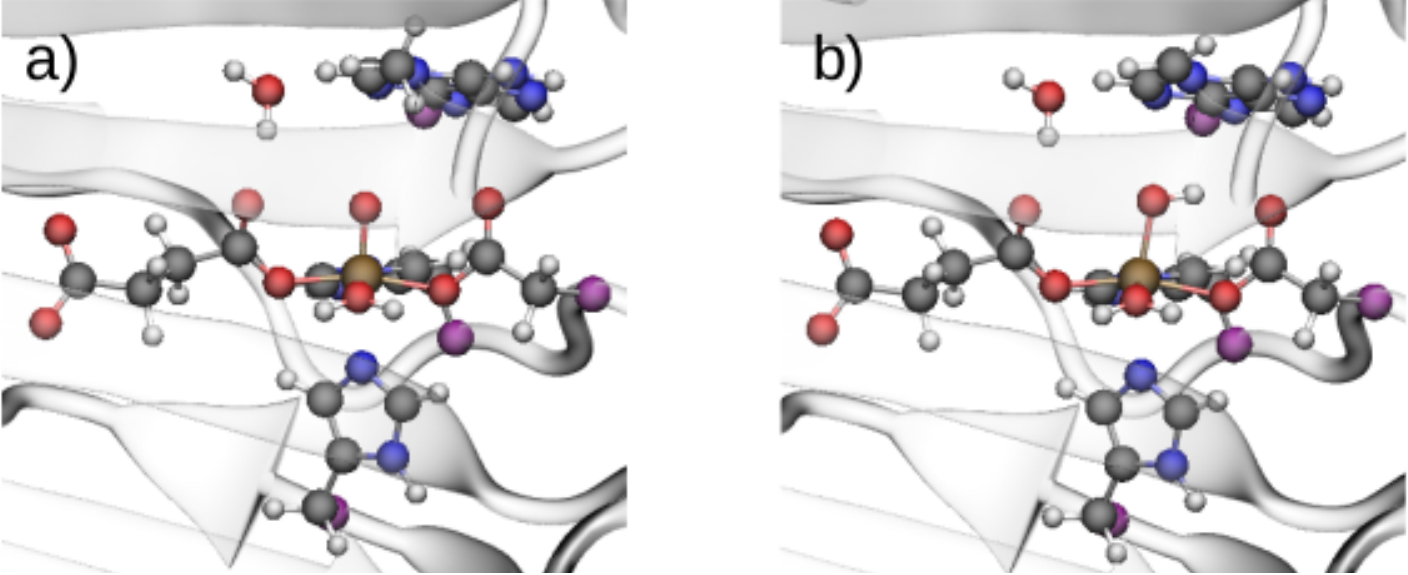}
 \includegraphics[height=5cm]{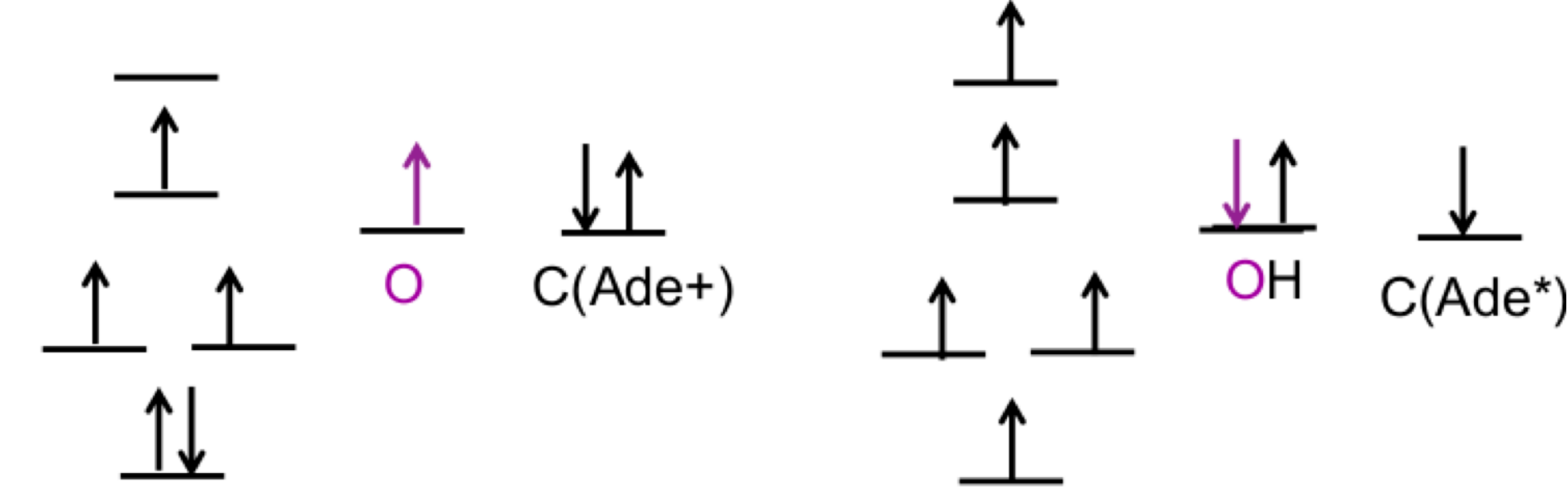}
 \caption{Converged reactant (a) and product (b) structures with their respective spin diagrams.} 
 \label{fig:ALKBH2-spin}
\end{figure*}

\subsection{Hydrogen atom abstraction}
Once the ferryl intermediate has been formed and the oxyl has
rearranged (\textbf{I4}, Figure \ref{fgr:alkb-mech}), 
the hydrogen atom transfer (HAT) step can take place
from the methyl on the substrate. This is the rate 
limiting step for the entire reaction (\textbf{I4} to \textbf{I5}). 
For this step, an electron must be transferred
into an orbital of the ferryl group. Two possible channels,
termed $\sigma$ and $\pi$ have been postulated depending on the
orbital that is occupied after the HAT.\cite{Geng105717}
In the $\sigma$-channel, a spin-up electron is transferred from the $\sigma$ 
orbital of the methylated substrate to the $\sigma^*$ orbital of the Fe(IV)=O. 
In the $\pi$-channel, a spin-down electron 
is transferred to the $\pi^*$ orbital of the Fe(IV)=O. These two 
channels arise due to the possible angles that can be 
formed by the Fe-O-H atoms in the TS, which lead to differing 
orbital overlaps. 
The TS structure of the $\sigma$-channel has a linear geometry ($\sim$180$^{\circ}$) 
while the Fe-O-H angle for the $\pi$-channel is significantly smaller
($\sim$120$^{\circ}$).\cite{Waheed20795} 

Calculations for the HAT step have been performed for several spin states including
triplet, quintet and septet, with the quintet 
state being the most stable consistent with previous work 
with TauD.\cite{Price037497,Bollinger054245} 
This  intermediate can adopt two spin states:  
an intermediate spin state that is ferromagnetically
coupled  to the oxyl group ($^{IS}$Fe--O$_F$), and 
a high spin Fe that is antiferromagnetically coupled to the 
oxyl group ($^{HS}$Fe--O\textsubscript{AF}). \cite{Fang136410}
The resulting \textbf{I4} state (Figure \ref{fgr:alkb-mech})
corresponds to the $^{IS}$Fe--O$_F$ state, with spin
density populations of 3.26 au on the Fe and 0.54 au on the O.
These spin densities are consistent with other reports on AlkB 
and other Fe=O systems (see below). \cite{Cisneros1070,Quesne14435,Fang136410}

Prior to the TS, an intersystem crossing (ISC) between the 
$^{IS}$Fe--O$_F$ and the $^{HS}$Fe--O\textsubscript{AF} states was 
observed.\cite{Fang136410} After the minimum energy
crossing point (MECP), the potential energy of the TS
structure for the hydrogen abstraction step corresponds to a 
barrier of 22.4 kcal/mol, with a calculated Helmholtz
free energy barrier of 18.9 kcal/mol. \cite{Fang136410} Similar 
results for potential energy barriers were 
reported  by Liu \textit{et al.} (20.9 kcal/mol), Quesne
\textit{et al.} (23.4 kcal/mol), and Waheed \textit{et al.}
\cite{Liu094887,Quesne14435,Waheed20795}
Interestingly, several of the reported works show that the 
chosen level of theory has a significant impact on the calculated
electronic structure description, with GGA functionals only
reporting one surface ($^{IS}$Fe--O$_F$) whereas simulations
that employ range--separated and dispersion corrected 
functionals report both spin surfaces. \cite{Fang136410,Liu094887,Quesne14435,Waheed20795}
These results are all in good agreement with the reported
barrier of 19.8 kcal/mol estimated from 
experimental kinetic analysis.\cite{Koivisto0344348}

\begin{table*}
\caption{Comparison of reaction energies and potential energy barriers for ALKBH2 and ALKBH3
compared to AlkB and to estimations from experimental kinetics data.}
\centering  
\begin{tabular}{ l c c }
  \hline
 & $\Delta$E$_{reac}$ (kcal/mol) & $\Delta$E$^{\ddagger}$(kcal/mol) \\
   \hline
AlkB\cite{Fang136410} & -3.7 & 22.4 (~19.8 exp.\cite{Koivisto0344348})\\
ALKBH2 & -3.2 & 25.7 (~21.3 exp.\cite{Ringvoll084142})\\
ALKBH3 & -4.2 & 28.6 \\ 
   \hline
\end{tabular}  
\label{tab:Energies}
\end{table*}

\begin{figure*}
    \centering
    \includegraphics[height=6.5cm]{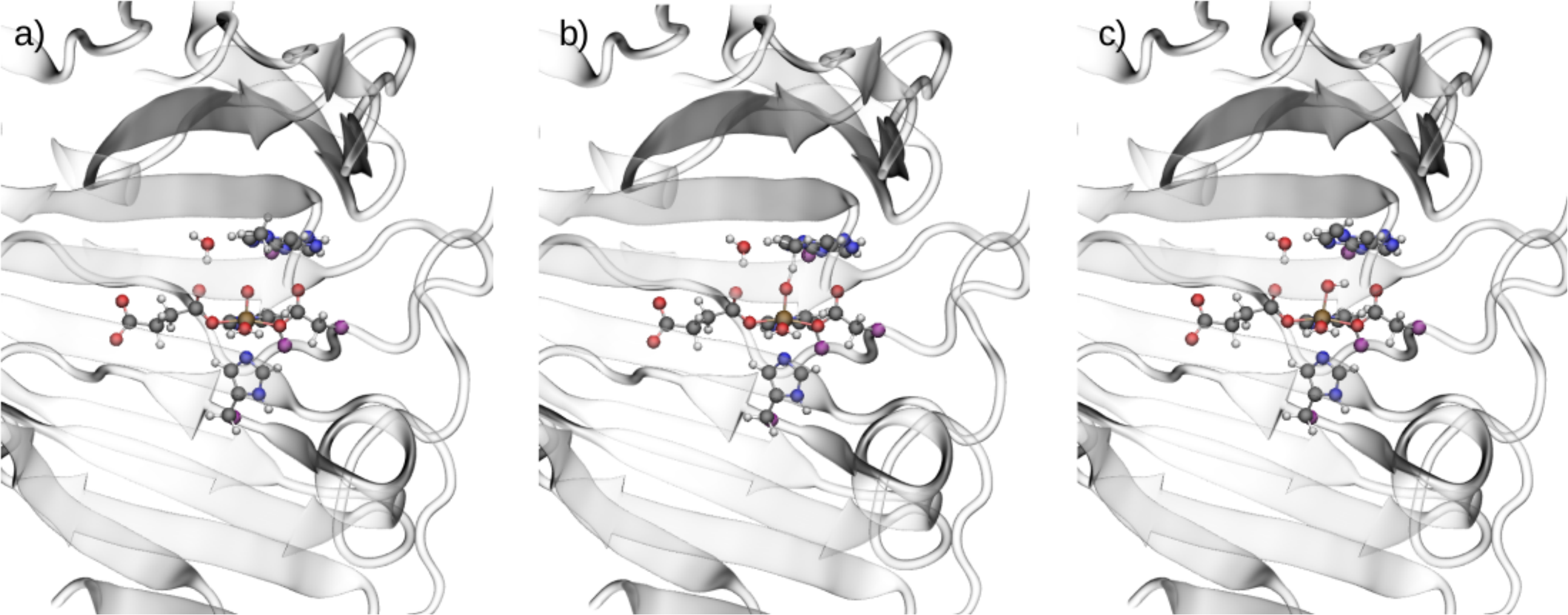}
    \includegraphics[height=6.5cm]{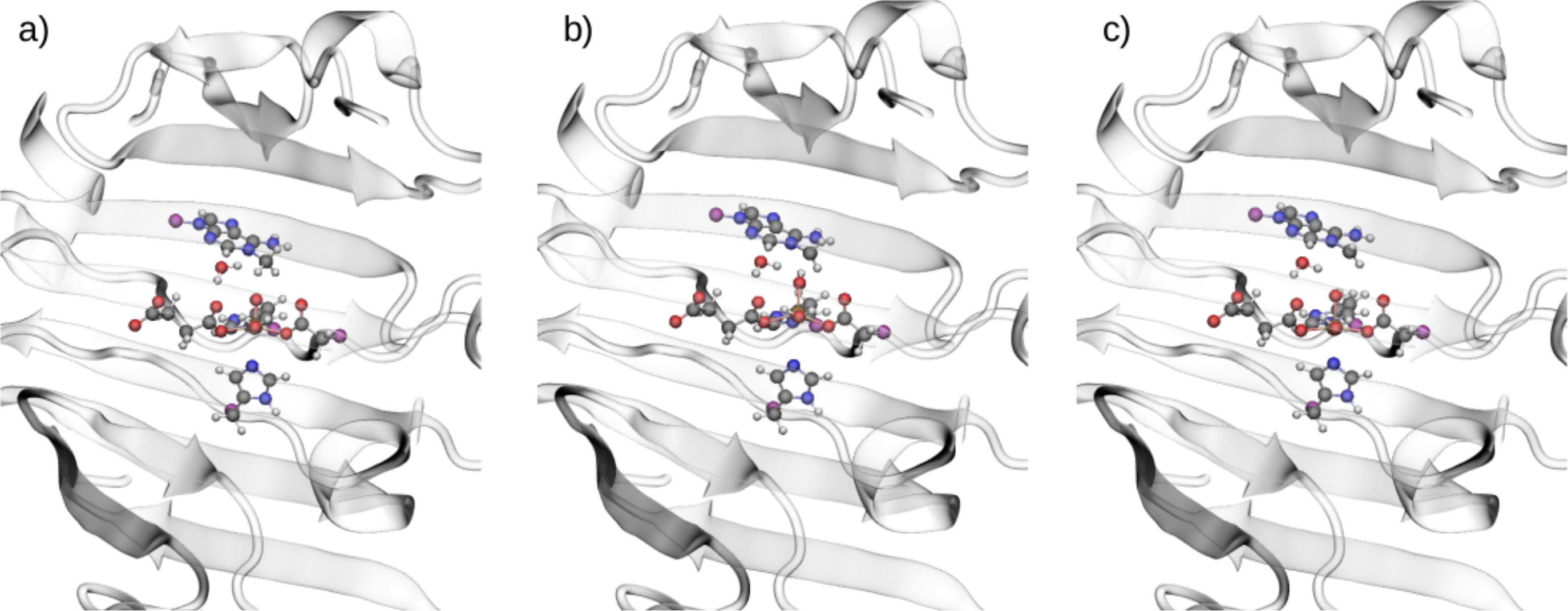}
    \caption{Optimized reactant (a), TS (b) and product (c) structures for ALKBH2 (top) 
    and ALKBH3 (bottom).}
    \label{fig:ALKBH2-RTSP} 
\end{figure*}

\subsubsection{QM/MM Simulations of \textbf{I4}$\rightarrow$\textbf{I5} by ALKBH2 and ALKBH3}
Humans have nine homologs of AlkB, ALKBH1-8 and FTO, that catalyze the repair 
of different DNA substrates. ALKBH2 and ALKBH3 can repair 
double-stranded DNA with a variety of lesions
including 3-methylcytosine (3mC), 1-methyladenine (1mA), 
3-methylthymidine (3mT), 1-methylguanine (1mG), and several 
etheno adducts. 
\cite{Bian195522,Fedeles1520734,Delaney0414051,Li131182}
ALKBH2 has a unique pincer structure to hold the 
double stranded DNA substrate in place, and has a very low 
efficiency for single stranded DNA substrates.\cite{Aas03859}
Contrary to ALKBH2, ALKBH3 has been shown experimentally to prefer ssDNA and act 
efficiently on several methylated RNA bases.\cite{Aas03859,Dango12373}

We performed QM/MM simulations  to investigate the reaction 
mechanism for the \textbf{I4} to \textbf{I5} step of the dealkylation
of 1mA catalyzed by ALKBH2 and ALKBH3 following the same approach as
that reported by Fang {\em et al.} for AlkB described in the
previous subsection.\cite{Fang136410} Briefly,
crystal structures of the two enzymes with bound substrate
(ALKBH2: 3BUC and ALKBH3: 2IUW) were used as starting points and 
examined with Molprobity followed
by system preparation including solvation in TIP3P water and neutralization of the
system with Na. \cite{Yang08961,Sundheim063389,Williams18293,Jorgensen83926,Schafmeister95} 
Subsequently, MD simulations were performed following
the same procedure as described by Fang \textit{et al.} 
using AMBER12's pmemd.cuda program and the ff99SB force
field. \cite{Fang136410,Case12,Lindorff-Larsen121} Each system
was subjected to 50 ns of production in the NVT ensemble with the 
Berendsen thermostat and a 1 fs timestep. \cite{Berendsen843684} 
Smooth particle mesh Ewald was used for the nonbonded interactions with an 8 \AA 
cutoff from which selected snapshots
were obtained for the subsequent QM/MM simulations (see SI for further computational methods).\cite{Kuwajima883751,Darden93}

\begin{table*}
\small
    \caption{Comparison of electronic and structural parameters for the rate-limiting step in
    AlkB\cite{Fang136410}, ALKBH2 and ALKBH3.}
    \label{tab:Parameters-Fe-oxo}
    \centering
    \begin{tabular}{llllllllll}
    \hline
    \multicolumn{1}{c}{{\textbf{Parameter}}} & \multicolumn{3}{c}{\textbf{AlkB}} & \multicolumn{3}{c}{\textbf{ALKBH2}} & \multicolumn{3}{c}{\textbf{ALKBH3}} \\
    \multicolumn{1}{c}{}       & R         & TS        & P         & R          & TS         & P         & R          & TS         & P         \\ \hline
    \textbf{Fe Spin Density}    & 3.26      & 4.24     & 4.35     & 3.20       & 4.16       & 4.36      & 3.31       & 4.33       & 4.34      \\ 
    \textbf{O Spin Density}     & 0.54      & -0.23    & 0.26     & 0.62       & -0.06      & 0.27      & 0.46       & 0.17       & 0.29      \\ 
   \textbf{Fe-O Distance (\AA)} & 1.61      & 1.77     & 1.84     & 1.60       & 1.74       & 1.84      & 1.59       & 1.86       & 1.83      \\  \hline
    \end{tabular}
\end{table*}

All QM/MM simulations for both systems involve QM regions containing 
70 atoms including 4 pseudobond atoms.
\cite{Parks08154106} The $\omega$B97XD functional was used with a 6-31G(d,p)
basis set for all non-pseudobond atoms.\cite{Fang136410} An in-house software using TINKER for the
MM part and a modified version of Gaussian16 for the QM part was used to
perform additive QM/MM with electrostatic embedding for ALKBH2.\cite{Tinker8,Gaussian16} 
The ALKBH3 results were obtained with the same
parameters and procedure using the LICHEM 1.1 software package.\cite{Kratz161019, Gokcan193056} 
Each converged structure 
was confirmed to have no negative frequencies except for the TS, 
which has one negative frequency along the reaction coordinate. 
The quadratic string method (QSM) was used to optimize the path between 
reactant, a guess TS and product for ALKBH3.\cite{Gokcan193056}
The MM environment was replicated from the reactant structure 
for all images, and initially restrained. These
restraints were gradually lowered over 11 steps.\cite{Gokcan193056} 

Figure \ref{fig:ALKBH2-RTSP} shows the optimized structures for the
critical points (\textbf{I4}, TS and \textbf{I5}) for ALKBH2 and
ALKBH3. 
Similar to what was reported by Fang \textit{et al.}, the \textbf{I4}
systems correspond to a $^{IS}$Fe--O$_F$ state in both cases.
Figure \ref{fig:ALKBH2-spin}a presents a diagram for the spin states
for the reactant and product on the quintet surface for ALKBH2. The minimized reactant 
has a calculated Mulliken spin density on the Fe of 3.20(3.31) for ALKBH2(ALKBH3), and the 
oxo has a spin density of 0.62(0.46) (\ref{tab:Parameters-Fe-oxo}), as can 
also be seen in Table \ref{tab:Parameters-Fe-oxo}. Additionally, 
these spin densities are comparable to those obtained for an intermediate spin 
Fe-O inorganic complex.\cite{Verma155770}  The TS and product structures 
have Fe spin densities of 4.16(4.35) and 4.36(4.34) for ALKBH2(ALKBH3) 
and O spin densities of -0.06(0.17) and 
0.27(0.29) au respectively. These values again 
compare quite well to previous work and literature values.\cite{Fang136410,Quesne14435,Waheed20795} 
The observed changes in spin 
density (Table \ref{tab:Parameters-Fe-oxo})
are in agreement with the work from Fang \textit{et. al.}, 
\cite{Fang136410} suggesting the existance of an ISC between 
\textbf{I4} and \textbf{I5} for these systems.

\begin{figure}
 \centering
 \includegraphics[height=5cm]{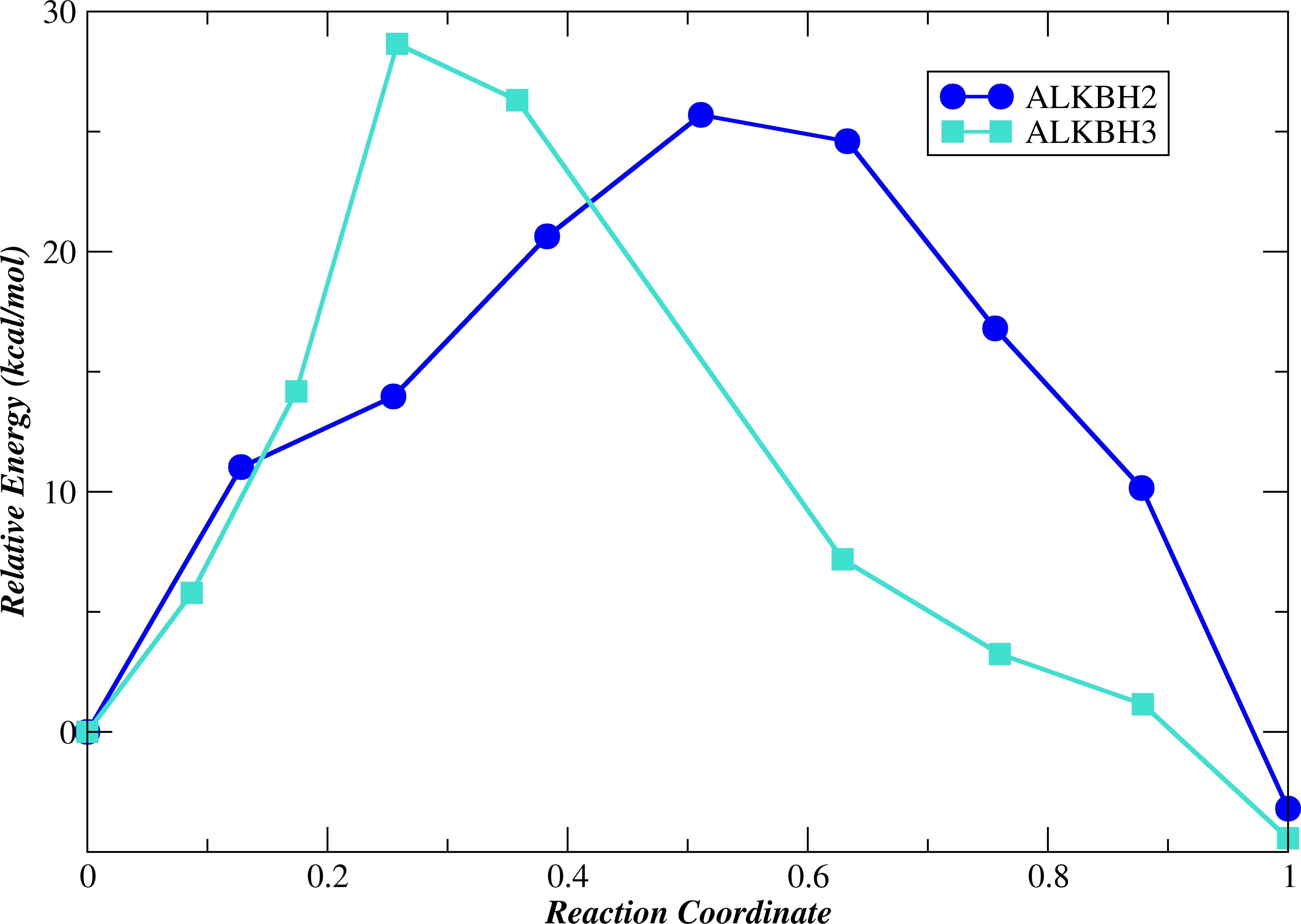}
 \caption{ALKBH2 and ALKBH3 optimized paths connecting reactant and product 
 structures using the quadratic string method.}
 \label{fig:ALKBH-rxn-coord}
\end{figure}

For ALKBH2, the calculated $\Delta E$ barrier is 25.7 kcal/mol, which compares favorably 
to the experimental $\Delta G$ of $\sim$ 21.3 kcal/mol as estimated from 
experimental kinetics data.\cite{Ringvoll084142} In the case of AlkBH3, the calculated potential
energy barrier is 28.6 kcal/mol. ALKBH2 has been shown to have lower processivity than 
AlkB for 1mA repair, and interestingly this point is captured by the calculated results 
(Table \ref{tab:Energies}). 

\subsection{Hydroxyl radical rebound}
\begin{figure*}
 \centering
 \includegraphics[height=5.5cm]{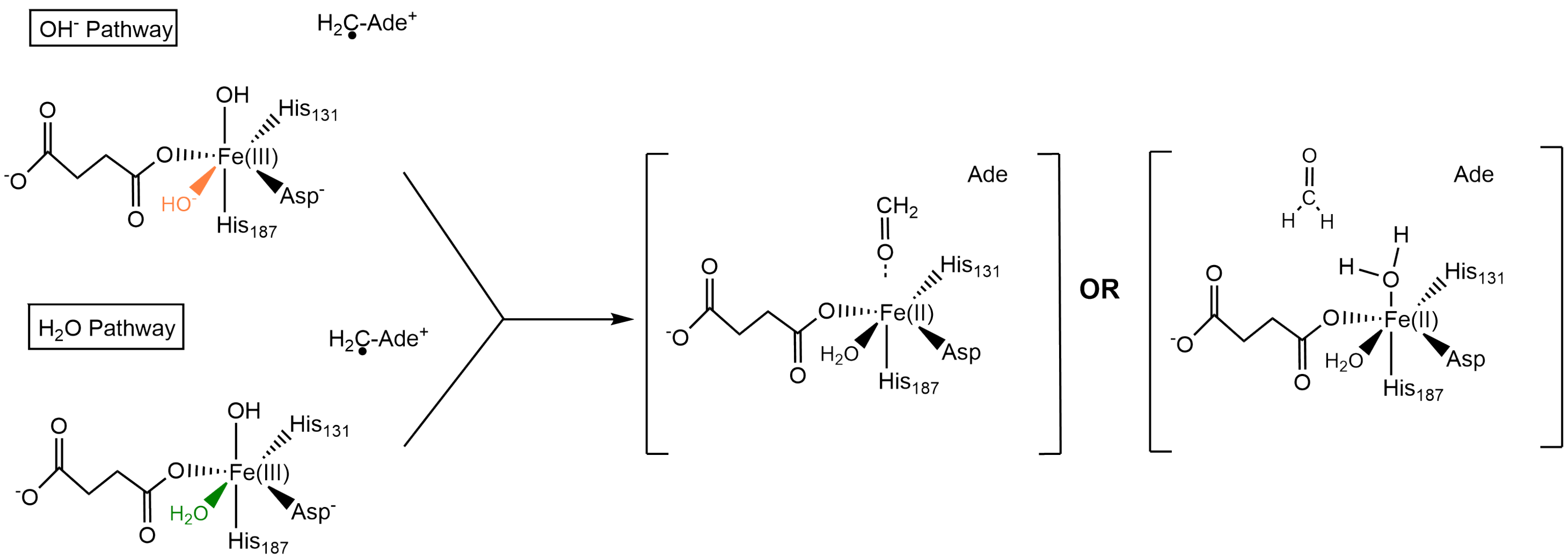}
 \caption{Two possible pathways following the hydrogen atom abstraction of AlkB.\cite{Fang145136}}
 \label{fgr:alkb-ohvsh2o}
\end{figure*}

The final phase of the reaction by AlkB involves the rebound 
of OH to the carbon radical of 1mA (\textbf{I5}). 
Finally, the C-N bond breaks and formaldehyde is formed (\textbf{I6}). Results 
from both crystal structure analysis
and time-resolved Raman suggest that a zwitterion intermediate may be possible.
\cite{Grzyska103982,Yi10330} 
In addition, there is a large vacancy (shown in Figure \ref{fgr:alkb-active-site}) 
along the plane that 
the succinate and Asp share. This could easily be occupied by a water molecule resulting in a 
H\textsubscript{2}O pathway or a hydroxide radical if the pH is slightly basic, resulting 
in a OH\textsuperscript{-} pathway (Figure \ref{fgr:alkb-ohvsh2o}). \cite{Fang145136} 
The barrier for the abstraction of a hydrogen atom is very similar for both pathways, 
23.2 kcal/mol (OH\textsuperscript{-} pathway) and 22.4 kcal/mol (H\textsubscript{2}O pathway).
\cite{Fang145136} While the barriers may be similar, the respective mechanisms  differ in 
significant ways. The rebound of the OH (\textbf{I6}) can occur via a concerted or stepwise path.
The concerted path involves the coupled OH rebound and proton transfer, 
whereas in the stepwise mechanism
the proton is first transferred followed by the oxygen. A concerted path was 
preferred in the OH\textsuperscript{-} path in which the OH rebounds and proton transfers
simultaneously followed by the breaking of the C-N bond.
The H\textsubscript{2}O pathway favors a sequential mechanism in 
which the hydroxyl unbinds from the Fe and 
then loses its proton. The loss of the proton is paired with the C-N bond breaking.
Several proton acceptors were proposed and analyzed including 
Asp133, Asp135, Glu136 or solvent. While the proton transfer to Glu136 had the lowest barrier,
the authors could not rule out the other options. The entire study suggested 
that because the barrier for the C-N bond breakage was smaller in the OH\textsuperscript{-}
and exterior proton acceptors are not needed, the OH\textsuperscript{-} is likely the
preferred pathway.\cite{Fang145136}

\subsection{Effects of second-sphere residues and other molecules} 
The environment surrounding the active site plays an important 
role in catalysis. As mentioned above, a second--shell water 
molecule that coordinates the ferryl moiety was observed to be
important for the rate--limiting step in AlkB, ALKBH2 and ALKBH3.
Several residues around the active site of AlkB are also known
to play a crucial role in catalysis.
In the case of AlkB, an energy decomposition analysis (EDA) 
was carried out by Fang \textit{et al.}, which 
can provide insight into the Coulomb and vdW interactions calculated during MD.
\cite{Fang136410} This analysis
suggested that nine residues exhibited energy differences that 
significantly impacted the reaction barrier: T51, R73, Y76, K127, Q132, Q155, R161, F185, and R210.
Only three of the nine have been studied experimentally (Y76, T51 and R161). R161
does not appear to have a significant effect on the reaction 
barrier, but does aid in stabilization of 1mA. Both T51 and Y76
contribute to a decrease in the reaction rate. These two sites were subjected to mutation to
an alanine residue, \cite{Holland10} which lead to an increase in the reaction barrier of 
0.5 kcal/mol (T51) and 1.0 kcal/mol (Y76). A sequence alignment was also carried out that compared
AlkB with ALKBH2 and ALKBH3, human homologues of AlkB. Two residues were structurally conserved:
R210 and R161.

Waheed \textit{et al.} used a hydrogen bond analysis to better understand the importance
of the residues surrounding the active site or QM region. The TS of AlkB was 
found to be stabilized by a $\pi$-stacking interaction between 
Y76 and W69 and the substrate's cytosine
ring. The Fe specifically was stabilized by a hydrogen bond interaction via R210. This 
same residue was proposed by Fang \textit{et al.} to stabilize the TS.\cite{Fang136410} 
In other work, it has also been suggested that R210 aids in 
hydrogen bonding with the oxo and succinate group. \cite{Quesne14435}
This aids in the isomerization necessary for the "in-line" binding mode of O\textsubscript{2}.
Lastly, I143 and W178 aid in the stabilization of the TS in which CO\textsubscript{2} is being lost
and the Fe reorients to an Fe(IV)=O intermediate. \cite{Waheed20795} The same site, W178, was also 
important for modulating the two possible tunnels for O\textsubscript{2} transport.
\cite{Torabifard176230}

\section{Mechanistic Studies of TET2}

\begin{figure}
 \centering
 \includegraphics[height=4.5cm]{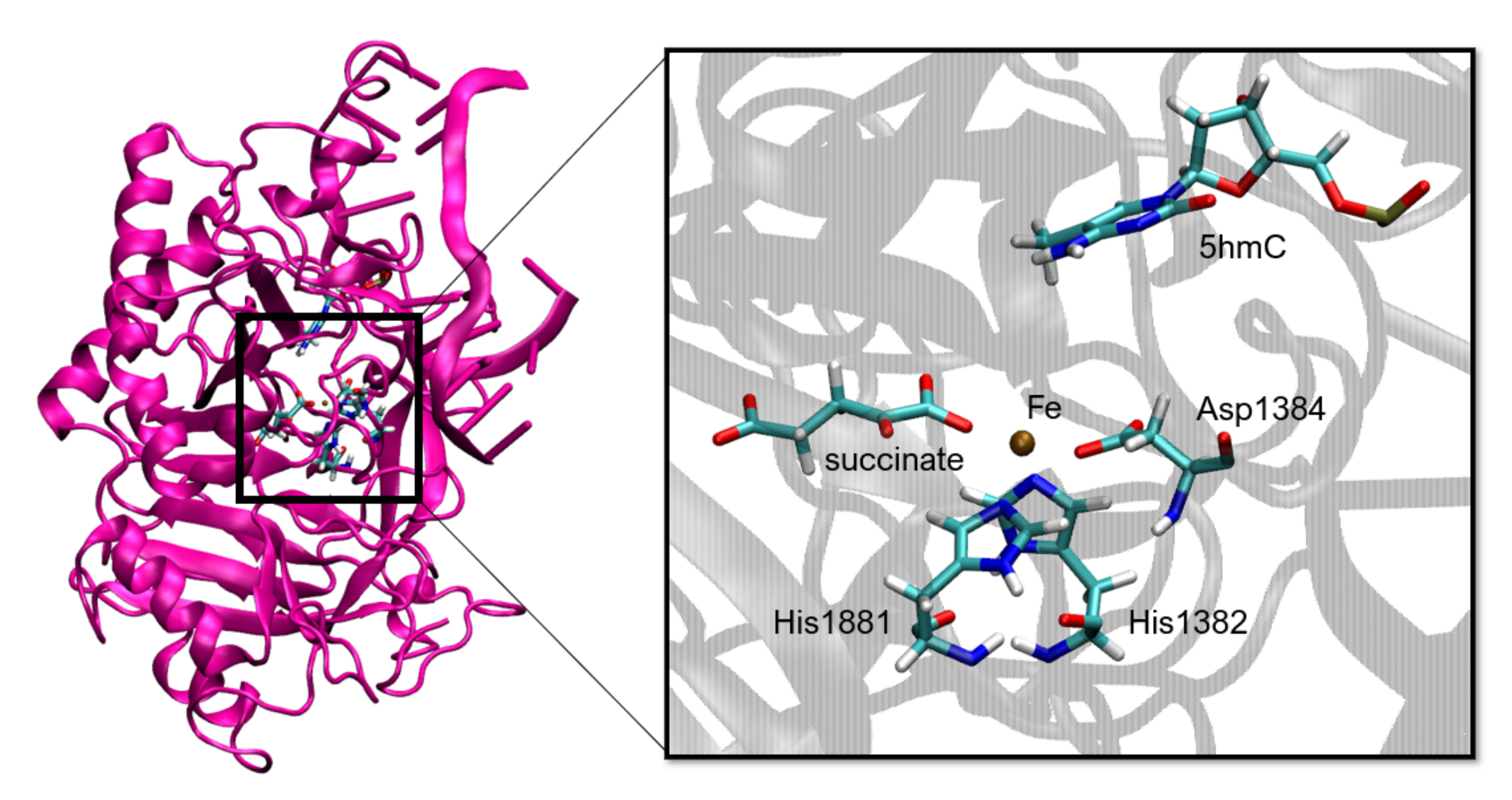}
 \caption{Crystal structure and active site of TET2 with succinate.\cite{Hu131545}}
 \label{fgr:tet2-active-site}
\end{figure}

\begin{figure}
 \centering
 \includegraphics[height=8.5cm]{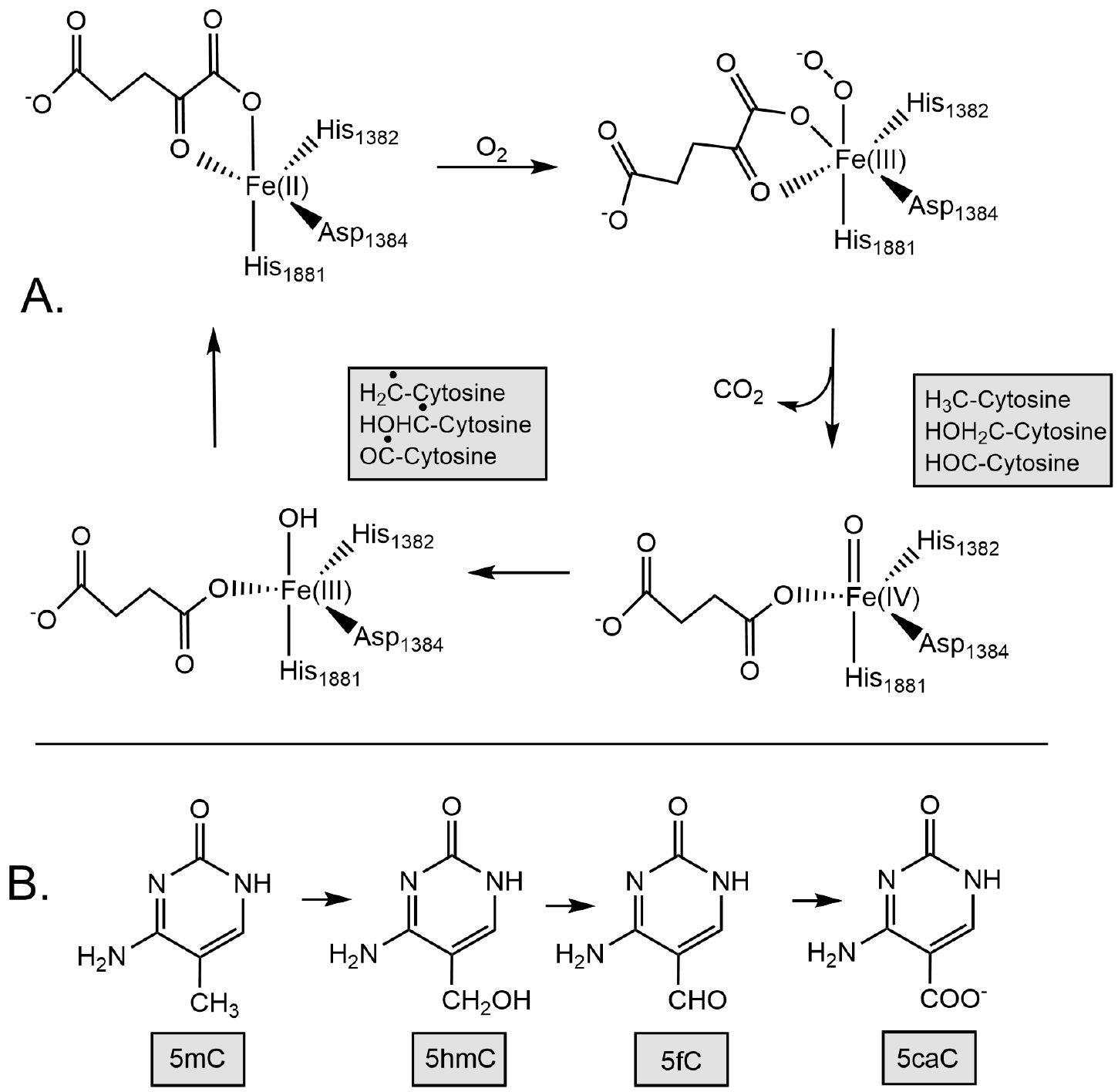}
 \caption{(A) The demethylation reaction carried out by TET2 and (B) the sequence of 
 methylated cytosine bases resulting from one round of oxidation.}
 \label{fgr:tet2-mech}
\end{figure}

TET2 (Figure \ref{fgr:tet2-active-site}), similar to AlkB, 
catalyzes the demethylation of DNA, albeit with
a major preference for a single substrate, 5mC, and
using an iterative oxidation mechanism (Figure \ref{fgr:tet2-mech}A).
Rather than a direct demethylation of 5mC, TET2 does so 
indirectly in three
sequential steps  following single-hit kinetics: 5mC to 5hmC, 5hmC to 5fC, and 
5fC to 5caC (Figure \ref{fgr:tet2-mech}B). 
\cite{Ito111300,Tahiliana09930,He111303,Hu15118}
Recently, it has been shown that TET1-3 enzymes can also carry out 
direct demethylation with a preference for substrates that lack a 5-methyl group
such as 4-N-methyl-5-methylcytosine or 4,4-N,N-dimethyl-5-methylcytosine. \cite{Ghanty2011312}

The indirect demethylation of each intermediate is similar to 
the AlkB enzyme mechanism in 
that these three phases are present: (1) formation of the reactive
Fe(IV)=O, (2)  hydrogen atom abstraction from the substrate, 
followed by (3) hydroxyl rebound (for 5hmC and 5caC) or second
H abstraction (for 5fC)  (Fig. \ref{fgr:tet2-mech}A). 
To determine if the hydrogen atom abstraction step was also the rate-limiting
step of TET2, Hu \textit{et al.} carried out a deuterium exchange analysis.\cite{Hu15118}
Hydrogen atoms on the methylated base (5mC) were replaced by
deuterium, which lead to a significant decrease in the enzymatic 
activity of TET2 thus confirming this is also
the rate-limiting step for TET2. 

\subsection{5mC to 5hmC}

The first stage in the TET2 catalyzed demethylation 
involves the oxidation of 5mC of 5hmC. This step is similar
to the AlkB catalyzed reaction, however, the main difference
is that there is no disproportionation to formaldehyde and cytosine
after the OH rebound step. It has been recently reported that
Alkb, ALKBH2 and ALKBH3 can also carry out the iterative
oxidation of 5mC, albeit with a 5--fold lower efficiency
compared to the repair of 1mA. \cite{Bian195522} Interestingly,
AlkB family enzymes can also repair other neutral alkylated bases
such as 3mT and 1mG. \cite{Delaney0414051,Koivisto0440470} However,
similar to the reduced activity for 5mC oxidation by AlkB enzymes,
the efficiency for the repair of these uncharged damaged bases
is also significantly reduced compared with charged substrates
repaired by AlkB family enzymes. \cite{Delaney0414051,Koivisto0440470}

Lu \textit{et al.} performed QM/MM simulations to investigate
all the oxidation steps catalyzed by TET2.\cite{Lu164728}
In the crystal structure of TET2 with 5mC used by 
Lu \textit{et al.} (PDB:4NM6), it was noted that residue D1384
adopted an orientation that was different when compared with the 
other two crystals in which 5hmC and
5fC were bound.\cite{Hu131545,Lu164728} In addition, no second-shell 
water was observed between D1384 
and 5mC, which they hypothesized to be consistent with the more 
hydrophobic methyl of 5mC. 

In the first reaction phase where the dioxygen is bound to 
Fe and attacks the $\alpha$-kg, this reaction
was found to be highly exergonic with a relatively small reaction
energy barrier of 7.6 kcal/mol for 5mC. The same reorientation
from axial to equatorial as seen in AlkB was also observed in 
TET2 for the oxidation of 5mC. Here, the 
dioxygen is initially trans to H1382 but reorients to trans w.r.t.
H1881. This reorientation was deemed necessary because of the large 
electrostatic attraction between R1286 and the peroxy
bridge. From there, the peroxy bridge must be broken 
leading to the formation of Fe(IV)=O. This phase 
of the reaction was found to be irreversible and quick to proceed. 

Once the Fe(IV)=O has formed, the rate-limiting step then proceeds. Previous
work has shown that for a quintet spin state, the $\sigma$-pathway is favored resulting in a linear
attack.\cite{Ye111228} This was also seen in the calculations by Lu \textit{et al.} and Waheed 
\textit{et al.}\cite{Lu164728,Waheed213877}
The relative potential energy barrier for this step was found to be 15.5 kcal/mol for
5mC.\cite{Lu164728}
Similar results were found for this step by Waheed \textit{et al.} (16.3
kcal/mol).\cite{Waheed213877}
The final stage of the reaction for this oxidation step
is the hydroxyl rebound.  Lu \textit{et al.} found a barrier of 16.9
kcal/mol, which suggests the hydroxyl rebound is slightly higher in energy 
than the HAT from the 5mC substrate.\cite{Lu164728} This effect 
is proposed to be due to the freely rotating methyl that allows for a low 
hydrogen abstraction barrier that allows for a hydrogen atom to 
be consistently oriented toward the activated
ferryl leading to an efficient abstraction.\cite{Lu164728} Conversely,
Waheed \textit{et al.} reported a barrier of 10.1 kcal/mol for the rebound 
step.\cite{Waheed213877}

\subsection{5hmC to 5fC}

The oxidation of 5hmC to 5fC involves a hydrogen atom transfer from
the hydroxyl moiety, and a proton transfer from the methylene moiety on
5hmC. Two separate groups have reported that the ferryl intermediate includes
a second--shell water that hydrogen bonds to the oxygen atom on the ferryl
intermediate. \cite{Hu15118,Lu164728,Torabifard188433}
Similar to the second--shell water for the ferryl intermediate in the 
AlkB family enzymes mentioned above, this water plays an important structural
and electronic role. \cite{Hu15118,Lu164728,Torabifard188433,Hix21} 
 
In the work performed by Lu \textit{et al.}, the reaction 
begins with the
formation of the peroxy bridge. This causes the hydroxyl group of 
5hmC to be in close enough proximity to interact with R1261 
and the peroxy moiety.\cite{Lu164728} In addition, experimental work 
has shown that 5hmC (and 5fC) is highly likely to form 
intra-molecular hydrogen bonds making the reaction more difficult to
proceed.\cite{Hu15118}
These additional interactions prevent the total reorientation of
the peroxy bridge and ultimately affect the geometry around the Fe 
center. \cite{Lu164728} In addition, the following phase of the reaction in 
which the 
bridge is broken leads to a square pyramidal geometry around the Fe 
as seen in 5mC and 5fC. However, the oxo group is trans w.r.t. H1382 rather 
than trans w.r.t H1882 as seen in the other two systems. Due to the 
bond between the oxo group and the water, the reorientation is crucial for 
the reaction to proceed to the rate-limiting step much like AlkB. 
The barrier for this reorientation found by Lu
\textit{et al.} was 6.9 kcal/mol.\cite{Lu164728} After the reorientation,
a bond with the second-shell water molecule is also observed leading to an 
unfavorable orientation between the oxo and the hydroxyl hydrogen of 
5hmC. \cite{Lu164728}  Based on this structural arrangement, the
first step for the HAT/proton transfer reported by Lu \textit{et al.}
corresponds to the proton abstraction from the methylene, with a barrier 
of 18.7 kcal/mol, followed by the HAT from the hydroxyl
with a corresponding barrier of 7.0 kcal/mol.

By contrast, Torabifard \textit{et al.} used QM/MM calculations
to investigate the rate limiting steps only (HAT and proton 
abstraction). \cite{Torabifard188433} These simulations were based on
structures obtained from previous simulations that reported the 
importance of an active site scaffold in TET2. \cite{Liu17181} The
optimized ferryl intermediate and the energetic features of the 
reaction reported by Torabifard \textit{et al.} share several
similarities with the work of Lu \textit{et al.} including the
second--shell water forming a hydrogen bond with the ferryl oxygen.
However, the orientation of the 5hmC substrate in the optimized
intermediate is such that the HAT from the hydroxyl moiety is found
to occur first, with a barrier of 20.1 kcal/mol. This is followed 
by the proton transfer with a corresponding barrier of 
7.3 kcal/mol. \cite{Torabifard188433}. Thus, although the
calculated barriers agree in magnitude, the two reported mechanisms
differ with regards to the specific rate--limiting step involved
in this stage. This point underscores the importance of the 
arrangement of all components in the active site (see below)
and suggests that both mechanisms may be at play depending on 
the specific geometry of the system.

\subsection{5fC to 5caC}

Similar to the 5hmC system, the crystal structure of TET2 with 5fC 
contains a second shell water
between D1384 and 5fC (PDB:5D9Y).\cite{Hu15118} The barrier 
for the formation of the peroxy bridge was found 
to be 7.8 kcal/mol.\cite{Lu164728} Similar to the 
5mC system, following the formation
of the peroxy bridge with Fe, the dioxygen changes 
from trans w.r.t. H1382 to trans w.r.t. H1881.\cite{Lu164728}
While 5fC does not require an additional reorientation as seen 
in the 5hmC reaction, the barrier for the hydrogen
atom transfer is significantly higher than the other two steps 
(26.7 kcal/mol).\cite{Lu164728} The group hypothesized that
this large difference in energy may result from the restrained 
conformation of the formyl group. Because the $\sigma$-
pathway is favored, the need for a linear attack is crucial which 
is difficult for the rigid formyl group of 5fC. 
A higher energy barrier hydroxyl rebound was also determined 
in this system (20.7 kcal/mol).\cite{Lu164728}

\subsection{The effects of mutations on TET2 reactivity}

While it has been known that TET2 can perform the sequential
oxidation of 5mC, 
the question was raised as to why 5fC and 5caC are formed if 5hmC is stable in cells. 
\cite{Liu17181} Liu \textit{et al.} proposed that TET2 is 
specifically built to provide support
for these higher order oxidations and a scaffold in the active site 
is responsible for the sequential oxidation of these intermediates. 
Two key  residues within this scaffold were found to be conserved 
among many of the TET enzymes (T1372 and Y1902) 
and subjected to experimental and computational mutagenesis studies. \cite{Liu17181} 
When saturation mutagensis
was initially performed on T1372, three distinct phenotypes were 
observed: 1) WT-like, 2) 5hmC dominant, and 3) catalytically dead. 

Subsequent MD simulations were carried out on all four methylated
intermediates for WT several T1372 TET2 variants. \cite{Liu17181} 
The mutants were found to have different effects on inter-molecular 
interactions within the active site depending on the identity of the 
residue employed to replace T1371. Hydrogen bond and EDA further 
revealed a scaffold that consists of T1372-Y1902, which is necessary 
to orient the substrate in the active site for efficient catalysis.
Subsequent MD simulations on Y1902 predicted that a Y1902F variant 
would exhibit a 5hmC-dominant phenotype and a T1372A/Y1902F variant 
would rescue catalytic activity. Experimental mutagenesis confirmed 
these predictions and provided support for the importance of this 
scaffold in the active site of TET2.

Furthering this work, a computational investigation into the 
catalytic mechanism for the oxidation of 5hmC by the T1372E TET2 
variant was carried out to investigate the reason for the 
5hmC-dominant phenotype for this variant. \cite{Torabifard188433} 
It was found that the T1372E variant
has a different electron configuration for the ferryl intermediate
(\textsuperscript{HS}Fe-O\textsubscript{AF}) arising from an altered
orientation of the substrate compared to WT, which
has a similar configuration to AlkB, ALKBH2 and ALKBH3 
(\textsuperscript{IS}Fe-O\textsubscript{F}) (see above). 
The mutation in this scaffold eliminates a crucial hydrogen
bond between T1372 and Y1902 and instead, hydrogen bonds to 5hmC. 
This altered orientation results in a shift in the second-shell water
position where the oxygen is pointed toward the oxo moiety. These 
changes lead to a highly stabilized product where hydrogen bond
can then form after the HAT between the second-shell water and 
the newly formed water. By extension, the newly
formed water can then form a hydrogen bond with 5fC. 
The orientation of 5hmC changes within the active site resulting in 
a almost two-fold increase in the energy barrier for the oxidation 
of 5hmC to 5fC by T1372E TET2. These results provide further
support for the importance of the proper orientation of the substrate
in the active site of TET2 facilitated by
the active site scaffold.
 
Hu \textit{et al.} investigated a variety of mutants and their effect on 5mC oxidation.
\cite{Hu15118}
They found that the double mutant K1299E-S1303N decreases TET2 activity significantly. 
QM/MM calculations were carried out on this mutant by Waheed \textit{et al}. \cite{Waheed213877}
A much higher barrier (25.0 kcal/mol) 
was found when compared with the WT (16.3 kcal/mol)
for the hydrogen atom abstraction step
indicating that this double mutation decreases the rate of reaction. 
In addition, the K1299E-S1303N mutant
proceeds via the $\pi$-pathway rather than the $\sigma$-pathway as found in the WT. 
Torabifard \textit{et al.} found that K1299 has a significant impact on
catalysis by EDA, and showed that K1299 aids in the stabilization
of one of the transition states of the oxidation of 5hmC. \cite{Torabifard188433} 
Additional QM/MM was done on the S1290A-Y1295A, Y1902A and 
N1387A mutants and similar barriers were found. \cite{Waheed213877}
In addition, Hu \textit{et al.} studied the 
R1261G mutant and found that TET2 activity was again disrupted by this mutation. \cite{Hu15118} 
Torabifard \textit{et al.} found strong interactions between R1261 and 
$\alpha$-kg indicating a large stabilization effect for the TS. \cite{Torabifard188433} 
This same residue was found to form hydrogen bonds with the 
hydroxyl group of 5hmC rather than the oxo group bonding to 5hmC 
thus requiring a reorientation of the Fe(IV)=O
moiety.\cite{Lu164728} This required reorientation leads to 
an increase in the energy barrier for 5hmC.

As mentioned above, there is typically a decreasing 
concentration of the analogues of 5mC found within organisms
and this is likely due to the conformation of the methylated 
substrate within the active site. Sappa \textit{et al.}
proposed that if the active site were altered, a more efficient 
turnover of the 5mC analogues could be found.\cite{Sappa21}
The group began by replacing a variety of hydrophobic and polar 
residues with alanine that were near to the active site.
From there, the enzyme was exposed to 5mC and the concentrations 
of the intermediates (5hmC, 5fC and 5caC) were
measured. Several mutants (T1372A, T1393A and T1883A) showed a 
high preference for 5hmC (\textgreater 80\%) which was speculated
to occur because of the need for 5hmC to change orientation to 
further oxidation (5hmC-dominant phenotype).
In addition, they found that the V1395A mutant synthesized 5caC 
as the dominant intermediate with the highest yield (94\%).
This mutant, when compared with WT, was also more efficient at 
acting upon 5mC and 5fC. Similar mutations with small
amino acid side chains were implemented for residue V1395 and 
both V1395G and V1395S were able to successfully oxidize 5mC to 5caC.
The authors believe the active site of TET2 may be crowded 
and smaller side chains may aid in a higher catalytic efficiency
of TET2.\cite{Sappa21}

\subsection{Substrate preference}

While a majority of the work done this far on the TET2 enzyme 
has been with respect to DNA substrates, it 
is worth discussing the ability of TET enzymes to repair 
RNA methylated bases. The various forms of methylated
cytosine such as 5mC, 5hmC, 5fC and 5caC can also be 
found in different types of RNA (mRNA, tRNA, rRNA
and ncRNA).\cite{Motorin101415,Squires125023,Huang165495} 
While concentrations may vary across different
organisms, the need for repair enzymes is still crucial. 
Experimental and computational work done by DeNizio
\textit{et al.} looked at the preference of TET2 for 
ssDNA, ssRNA, dsDNA and dsRNA.\cite{DeNizio19411,LEDDIN201991} 
Their results indicate that while TET2 can tolerate 
both ssDNA and ssRNA, it prefers ssDNA which 
was also corroborated by Fu \textit{et al.}. \cite{Fu1411582} 
Similarly, dsDNA is highly preferred to
dsRNA and hybrid DNA:RNA configurations can also be 
tolerated if the substrate is DNA. It was speculated
that this strong disfavor for dsRNA may arise from the 
preference of RNA for the A-form structure.

\section{Summary and Perspective}

Computational methods paired with experimental work have 
shown to be very useful in elucidating the mechanistic
details of the Fe/$\alpha$-kg dependent enzymes discussed 
here. Both the AlkB and TET families are an integral
part of the DNA/RNA repair and modification enzymes 
found across all species and are particularly important for cancer prevention. 
The AlkB family still has some unanswered questions 
that require further work. The O\textsubscript{2} tunnels 
proposed by Torabifard \textit{et al.} indicate that 
O\textsubscript{2} could bind in two different positions
(trans w.r.t. H187 or trans w.r.t. H131) leading to different reaction paths. 
The binding of O\textsubscript{2} is a pivotal phase of the reaction
that will ultimately allow for the activation of the Fe moiety.

While both enzyme families can dealkylate a variety
of substrates, they each have preferences for certain bases. 
Mechanistically, AlkB can directly dealkylate damaged 
bases via one round of catalysis whereas TET2 can catalyze
both direct and indirect dealkylation.
The question can still be raised as to 
why these enzymes have specific preferences. 
AlkB, ALKBH2, ALKBH3 and TET2 have all been shown to 
demethylate 5mC and its derivatives.\cite{Bian195522}
Bian \textit{et al.} proposed that the AlkB enzymes need 
to bind 5mC in the syn-conformation to
allow the reaction to proceed. This binding conformation 
would be similar to the preferential methylated bases
that AlkB enzymes commonly repair such as 3mC. Additionally, the 
preferred TET2 substrate is neutral, while several AlkB family preferred
substrates are cationic, with possible dealkylation of neutral damaged bases. 
Nevertheless, the activity of AlkB family enzymes is significantly reduced
for neutral alkylated bases. Additionally, the role of second--shell
residues and other molecules has been shown to be important
for these enzymes, similar to other inorganic systems that carry
out similar C-H activation reactions \cite{Fang136410, Torabifard188433, Vitillo2021579}. 
These second--shell residues
are also likely to play an important role in selectivity.
Future work may help shed light on the drivers for the 
selectivity of members from both of these families.

\section*{Author Contributions}
Author contributions are as follow: ARW and EVM performed data curation, 
formal analysis, visualization and 
investigation for the ALKBH2 and ALKBH3 section. MBB performed 
data curation, investigation, formal analysis, 
visualization and writing of the original draft. GAC was involved in
supervision, funding securement, conceptualization, formal analysis, 
methodology and software. All authors 
contributed to reviewing and editing.

\section*{Funding}
This work was funded by NIH R01GM108583 and NSF CHE-1856162. 
Computing time from CASCaM with partial support from NSF CHE-1531468 is gratefully acknowledged.
MBB thanks MolSSI for support via a software development fellowship. 
E.A.V.-M. wishes to acknowledge CONACyT for funding.

\section*{Conflicts of interest}
Authors declare no conflicts.

\begin{acknowledgments}
The authors thank Prof. L. Noodleman and Prof. R. Lord for insightful discussions.
\end{acknowledgments}


\section{Appendix}
\section*{Methodology for the QM/MM ALKBH2 and ALKBH3 Calculations}
QM/MM single point calculations were performed for each critical point structure
(reactant, transition state and product) for ALKBH2 and ALKBH3.
The $\omega$B97XD functional, and the 6-31G(d,p) basis set were used for the atoms in
the QM region. The pseudo-bond
approach was used for the bonds that bridge the QM and MM regions.\cite{Parks08154106}
A modified version of the AMBER ff99SB force field including
the parameters of the damaged bases was used to model
the alkylated double and single stranded substrate and 
reactive site for ALKBH2 and ALKBH3 MM regions, respectively.
The TiP3P force field was used to model the MM water molecules. 
Gaussian16 and TINKER8 software packages were used with LICHEM
1.1.\cite{Gaussian16,Tinker8,Kratz161019,Gokcan193056} 

For the QM single point calculations, the $\omega$B97XD functional, 6-31G(d,p) basis set 
and the pseudo-bond approach 
for the cleaved bonds between the QM and the MM regions was used; each QM 
cluster was embedded in the electrostatic 
field produced by the MM region (ALKBH2/ALKBH3 + water). Mulliken spin densities 
for Fe were extracted from the 
Gaussian16 output file and the electron densities from the formatted check point 
file employing the Multiwfn 3.8 
software package.\cite{Multiwvfn}

\bibliography{references}

\end{document}